\documentclass[twocolumn]{aastex701}

\usepackage{amsmath}
\usepackage{booktabs}
\usepackage{array}
\usepackage{multirow}
\usepackage{graphicx}
\usepackage{siunitx}
\usepackage{physics}
\usepackage{amsbsy}

\makeatletter
\if@two@col
  \newcommand{\twocolswitch}[2]{#1}
\else
  \newcommand{\twocolswitch}[2]{#2}
\fi
\makeatother

\begin{document}

\title{Determining the Milky Way gravitational potential without selection functions}

\author[orcid=0000-0002-0417-8645]{Taavet Kalda}
\affiliation{Max-Planck-Institut f\"{u}r Astronomie,
K\"{o}nigstuhl 17, D-69117 Heidelberg, Germany}
\email[show]{kalda@mpia.de}

\author[orcid=0000-0001-5417-2260]{Gregory M. Green}
\affiliation{Max-Planck-Institut f\"{u}r Astronomie,
K\"{o}nigstuhl 17, D-69117 Heidelberg, Germany}
\affiliation{Westlake University, Department of Astronomy,
600 Dunyu Rd., Xihu District, Hangzhou, Zhejiang, People's Republic of China}
\email[show]{gregory@westlake.edu.cn}

\begin{abstract}

Selection effects, such as interstellar extinction and varying survey depth, complicate efforts to determine the gravitational potential -- and thus the distribution of baryonic and dark matter -- throughout the Milky Way galaxy using stellar kinematics. We present a new variant of the ``Deep Potential'' method of determining the gravitational potential from a snapshot of stellar positions and velocities that does not require any modeling of spatial selection functions. Instead of modeling the full six-dimensional phase-space distribution function $f\left(\vec{x},\vec{v}\right)$ of observed kinematic tracers, we model the conditional velocity distribution $p\left(\vec{v}\mid\vec{x}\right)$, which is unaffected by a purely spatial selection function. We simultaneously learn the gravitational potential $\Phi\left(\vec{x}\right)$ and the underlying spatial density of the entire tracer population $n\left(\vec{x}\right)$ -- including unobserved stars -- using the collisionless Boltzmann equation under the stationarity assumption. The advantage of this method is that unlike the spatial selection function, all of the quantities we model, $p\left(\vec{v}\mid\vec{x}\right)$, $\Phi\left(\vec{x}\right)$, and $n\left(\vec{x}\right)$, typically vary smoothly in both position and velocity. We demonstrate that this ``conditional'' Deep Potential method is able to accurately recover the gravitational potential in a mock dataset with a complex three-dimensional dust distribution that imprints fine angular structure on the selection function. Because we do not need to model the spatial selection function, our new method can effectively scale to large, complex datasets while using relatively few parameters, and is thus well-suited to \textit{Gaia} data.

\end{abstract}


\section{Introduction}
\label{sec:intro}

While the standard cosmological model involving cold dark matter (CDM) has been highly successful in predicting the large-scale structure of the Universe, the distribution of dark matter on sub-galactic scales, which is shaped both by baryonic feedback and the properties of the dark matter, remains less well understood \citep{BullockBoylanKolchin2017LambdaCDMSmallScaleChallenges}. As the only galaxy that we can study star by star, the Milky Way provides a unique opportunity to measure the distribution of dark matter on small scales. The gravitational potential is sourced by all matter; when combined with observationally informed models of the distribution of stars and gas, the potential thus contains information about the distribution of dark matter on small scales.

The most direct way of measuring the gravitational potential -- by observing the accelerations of stars orbiting in the Galaxy -- is at present beyond our technical reach in all but a limited number of cases \citep{SilverwoodEasther2019StellarAccelerationsGalacticGravitationalField, Chakrabarti2020TowardDirectMeasureGalacticAcceleration, Chakrabarti2022EclipseTimingMilkyWayGravitationalPotential, Moran2024PulsarBasedMapGalacticAcceleration}. A number of methods have therefore been developed to infer the gravitational potential from a snapshot of stellar kinematics, typically involving assumptions about the stationarity and symmetry (\textit{e.g.}, axisymmetry) of the Milky Way. Among these methods are Jeans modeling \citep{Jeans1915OnTheoryStarStreaming, Jeans1922MotionsOfStars, Binney1980JeansEllipticalGalaxies, Cappellari2008AnisotropicJeansModels}, Schwarzschild modeling \citep{Schwarzschild1979NumericalModelTriaxialStellarSystem, vandenBosch2008TriaxialOrbitBasedGalaxyModels}, and action-based modeling \citep{Binney2012ActionsAxisymmetricPotentials, Binney2012ActionBasedModelingGalaxy, BovyRix2013MilkyWayDiskSurfaceDensity, Trick2016ActionBasedModelingRoadMapping, Vasiliev2019AGAMAActionBasedGalaxyModelling}.

In recent years, the European Space Agency's \textit{Gaia} mission \citep{Prusti2016GaiaMission} has dramatically increased the quantity and quality of measured stellar parallaxes, proper motions, and radial velocities available. As of \textit{Gaia} Data Release~3, high-quality, six-dimensional phase-space measurements are available for $\sim$ 34~million stars \citep{Vallenari2023GaiaDR3}. The size of this dataset not only makes it possible -- but also makes it necessary -- to consider more flexible, nonparametric models of Galactic structure that do justice to the richness of the data.

\citet{GreenTing2020DeepPotential} and \citet{GreenTingKamdar2023DeepPotential} introduced a new method, ``Deep Potential,'' which makes minimal assumptions about the form of the gravitational potential and distribution function, using neural networks and normalizing flows \citep{TabakTurner2012NormalizingFlows,Papamakarios2019NormalizingFlowsReview} to flexibly model these functions. Deep Potential uses the collisionless Boltzmann equation (CBE) to find the gravitational potential that renders the observed distribution function as stationary as possible. Subsequent work has extended this method and applied it to both mock and real \textit{Gaia} datasets \citep{An2021UniquePotentialFromDF,Naik2022LearnAccelerationsSolarNeighbourhood, Buckley2023MeasuringDMUnsupervisedML, Kalda2024DeepPotentialRotatingFrame, Putney2024DeepPotentialDustCorrection, KaldaGreen2025DeepPotentialMW1kpc,Lim2025MappingDarkMatterGaiaDR3,Putney2025}.

In this work, we present a new variant of Deep Potential that obviates the need to model spatial selection functions, by learning the velocity distribution conditional on position, $p\left(\vec{v}\mid\vec{x}\right)$, from the observed kinematic tracers, and then learning the true spatial density of the full tracer population (including those that are unobserved due to selection effects), $n\left(\vec{x}\right)$, from the collisionless Boltzmann equation. We demonstrate that this method is able to accurately recover the gravitational potential on a mock dataset ``observed'' through a complex three-dimensional dust distribution. This method should alleviate a key challenge in applying Deep Potential to \textit{Gaia} data, and may even open up the possibility of applying the method to spectroscopic surveys with more patchy sky coverage and more complex selection functions, such as SDSS-V \citep{Kollmeier2017SDSSVPioneeringPanopticSpectroscopy}.

\section{Derivation of Method}
\label{sec:method}

\begin{table*}
  \centering
  \begin{tabular}{@{}lccl@{}}
    \hline
    \hline
           & \multicolumn{2}{c}{Functions learned from} \\
           \cline{2-3}
    Method & Observed tracers & CBE & Notes \\
    \hline
    ``Vanilla'' & $f\left(\vec{x},\vec{v}\right)$ & $\Phi\left(\vec{x}\right)$ & $S\left(\vec{x}\right)$ must be modeled based on prior knowledge. \\
    ``Selection function'' & $f_{\mathrm{obs}}\left(\vec{x},\vec{v}\right)$ & $\Phi\left(\vec{x}\right)$, $S\left(\vec{x}\right)$ & \\
    ``Conditional'' & $p\left(\vec{v}\mid\vec{x}\right)$ & $\Phi\left(\vec{x}\right)$, $n\left(\vec{x}\right)$ & $n_{\mathrm{obs}}\left(\vec{x}\right)$ can be optionally learned from the tracers. \\
    \hline
  \end{tabular}
  \caption{Overview of the three Deep Potential variants discussed in this work, indicating which functions are learned from the observed kinematic tracers (using normalizing flows) and from the collisionless Boltzmann equation (CBE; using neural networks).}
  \label{tab:methods}
\end{table*}

In the original (``vanilla'') formulation of Deep Potential \citep{GreenTing2020DeepPotential,GreenTingKamdar2023DeepPotential}, one fits a normalizing flow to represent the phase-space density $f\left(\vec{x},\vec{v}\right)$ of a population of kinematic tracers (\textit{i.e.}, stars) with observed positions $\vec{x}$ and velocities $\vec{v}$ (\textit{e.g.}, from \textit{Gaia}). One then calculates the gradients $\partial_{\vec{x}} f\left(\vec{x},\vec{v}\right)$ and $\partial_{\vec{v}} f\left(\vec{x},\vec{v}\right)$ at a set of phase-space coordinates drawn from the distribution function. Using the collisionless Boltzmann equation (CBE), and given a model of the gravitational potential $\Phi\left(\vec{x}\right)$, one can calculate the rate of change of the distribution function at any point in phase space:
\begin{align}
  \partial_t f\!\left(\vec{x},\vec{v}\right)
  &=
    \sum_{i=1}^3 \! \left[
      \frac{\partial \Phi\!\left(\vec{x}\right)}{\partial x_i}
      \frac{\partial}{\partial v_i}
      -
      v_i \frac{\partial}{\partial x_i}
    \right] \!
    f\!\left(\vec{x},\vec{v}\right)
  \, .
  \label{eqn:CBE}
\end{align}
Under the stationarity assumption, the left-hand side ($\partial_t f$) must vanish. Modeling the potential as a neural network (which takes a 3-vector representing $\vec{x}$ and outputs a scalar representing $\Phi$), one minimizes (with respect to the network parameters) a loss function that penalizes both non-stationarity ($\partial_t f \neq 0$) and negative matter densities ($\nabla^2\Phi < 0$):
\begin{align}
  L_{\mathrm{vanilla}} &=
    \Big\langle
      \sinh^{-1}\!\left(\alpha\left|\partial_t f\right|\right)
      \twocolswitch{\notag \\ & \hspace{1.0em}}{}
      +
      \lambda \sinh^{-1}\!\left(
        \beta\,\mathrm{max}\left\{-\nabla^2\Phi,0\right\}
      \right)
    \Big\rangle_{\vec{x},\vec{v} \sim f}
  \, ,
  \label{eqn:loss_vanilla}
\end{align}
where $\alpha$, $\beta$ and $\lambda$ (all set to unity throughout this work) are hyperparameters that control the shape and relative importance of the two penalties, and the average is taken over points in $\vec{x}$ and $\vec{v}$ drawn from the modeled distribution function. Minimizing $L_{\mathrm{vanilla}}$ yields a model of the gravitational potential $\Phi\left(\vec{x}\right)$ of the system (up to a physically meaningless additive constant).

This approach requires a fair sample of phase-mixed kinematic tracers. In the presence of dust extinction, varying survey depth, and other selection effects, one generally obtains a biased sample of the distribution function. To a large extent, these effects depend on where the star is located, not on how it moves. In principle, data processing pipeline effects, such as quality filtering in the case of \emph{Gaia}, could introduce weak, indirect velocity-dependent biases, since spectral quality correlates with stellar properties that, in turn, correlate with kinematics \citep{Katz2023}. In practice, however, these effects are expected to be small. To a good approximation, the selection function can thus be represented as a function of position, but not of velocity: $p\left(\mathrm{observed}\mid\vec{x},\vec{v}\right) = S\left(\vec{x}\right)$. We denote the distribution of observed kinematic tracers as

\begin{align}
  f_{\mathrm{obs}}\left(\vec{x},\vec{v}\right)
  = f\left(\vec{x},\vec{v}\right) S\left(\vec{x}\right)
  \, .
\end{align}
One approach to dealing with the selection function is to weight the observed tracers by $S^{-1}\left(\vec{x}\right)$ when learning the distribution function, in order to obtain an unbiased estimate of the true distribution function \citep{KaldaGreen2025DeepPotentialMW1kpc}. However, this approach requires an accurate estimate of the selection function $S\left(\vec{x}\right)$ at the location of each observed tracer, and amplifies noise in regions of low $S\left(\vec{x}\right)$.

\citet{Putney2024DeepPotentialDustCorrection, Putney2025} instead treats $\ln S\left(\vec{x}\right)$ as an additional function to be learned from the CBE, analogously to $\Phi\left(\vec{x}\right)$. Writing the CBE in terms of $\ln f\left(\vec{x},\vec{v}\right) = \ln f_{\mathrm{obs}}\left(\vec{x},\vec{v}\right) - \ln S\left(\vec{x}\right)$,
\begin{align}
  \partial_t \ln f\!\left(\vec{x},\vec{v}\right)
  &= \!
    \sum_{i=1}^3 \! \left\{
      \frac{\partial \Phi\!\left(\vec{x}\right)}{\partial x_i}
      \frac{\partial}{\partial v_i}
      \ln f_{\mathrm{obs}}\!\left(\vec{x},\vec{v}\right)
      \right.\notag\\&\left.-
      v_i \frac{\partial}{\partial x_i}
      \left[
        \ln f_{\mathrm{obs}}\!\left(\vec{x},\vec{v}\right)
        - \ln S\!\left(\vec{x}\right)
      \right]
    \right\}
  \, .
\end{align}
Because the terms involving the gravitational potential and selection function depend differently on velocity, it is possible to simultaneously learn both using nearly the same loss function as before (Eq.~\ref{eqn:loss_vanilla}). Replacing $\partial_t f$ with $\partial_t \ln f$ and adding a term that penalizes large deviations of $S\left(\vec{x}\right)$ from unity (with corresponding hyperparameter $\gamma_S$, which we set to zero in this work), one obtains the loss function:
\begin{align}
  L_{\mathrm{selfn}} &=
    \Big\langle
      \sinh^{-1}\!\left(\alpha\left|\partial_t \ln f\right|\right)
      \twocolswitch{\notag \\ & \hspace{1.0em}}{}
      +
      \lambda \sinh^{-1}\!\left(
        \beta\,\mathrm{max}\left\{-\nabla^2\Phi,0\right\}
      \right)
      \twocolswitch{\notag \\ & \hspace{1.0em}}{}
      +
      \gamma_S \left(\ln S\right)^2
    \Big\rangle_{\vec{x},\vec{v} \sim f_{\mathrm{obs}}}
  \, .
  \label{eqn:loss_selfn}
\end{align}
One attractive aspect of this ``selection function'' approach is that it obviates the need to determine the selection function beforehand. Additionally, if one has an initial estimate of the selection function (\textit{e.g.}, based on knowledge of the survey and dust properties), one can first learn an initial estimate of the distribution function using weighted kinematic tracers, and then treat $S\left(\vec{x}\right)$ as a learned correction to the selection function. However, one difficulty in this approach is that the selection function $S\left(\vec{x}\right)$ typically has highly complicated spatial structure (in particular, fine angular structure, as viewed from Earth, imprinted by dust extinction and varying survey depth), which is difficult for a neural network model to capture. The observed distribution function $f_{\mathrm{obs}}\left(\vec{x},\vec{v}\right)$ also typically contains detailed angular structure that even large normalizing flow models struggle to accurately capture.

We therefore propose a method that only requires measuring the distribution of velocities, conditional on position. As in \citet{Buckley2023MeasuringDMUnsupervisedML}, one can decompose the distribution function as $f\left(\vec{x},\vec{v}\right) = n\left(\vec{x}\right) p\left(\vec{v}\mid\vec{x}\right)$, where $n\left(\vec{x}\right)$ is the true spatial density of kinematic tracers (including those that are not observed due to selection effects). As the conditional velocity distribution, $p\left(\vec{v}\mid\vec{x}\right)$, is unaffected by a purely spatial selection function, it can be learned from the observed kinematic tracers using a conditional normalizing flow. The CBE can be rewritten in terms of $n\left(\vec{x}\right)$ and $p\left(\vec{v}\mid\vec{x}\right)$:
\begin{align}
  \partial_t \ln f\!\left(\vec{x},\vec{v}\right)
  &=
    \sum_{i=1}^3 \! \left\{
      \frac{\partial \Phi\!\left(\vec{x}\right)}{\partial x_i}
      \frac{\partial}{\partial v_i}
      \ln p\!\left(\vec{v}\mid\vec{x}\right)
      \twocolswitch{\right. \notag \\ & \hspace{1.5em} \left.}{}
      -
      v_i \frac{\partial}{\partial x_i}
      \left[
        \ln p\!\left(\vec{v}\mid\vec{x}\right)
        + \ln n\!\left(\vec{x}\right)
      \right]
    \right\}
  \, .
\end{align}
One can model the true spatial density of tracers, $\ln n\left(\vec{x}\right)$, as a neural network, and use the loss function Eq.~\eqref{eqn:loss_conditional} to simultaneously learn both $\ln n\left(\vec{x}\right)$ (up to an additive constant) and $\Phi\left(\vec{x}\right)$ from the data. The key to this ``conditional'' approach is that the model of the selection function is replaced by a model of the true spatial distribution of the tracers. While the former typically has fine spatial (particularly angular) structure, the latter does not.

While this method appears to discard the information contained in the spatial distribution of the observed tracers, it has a number of major advantages. First, the observed spatial distribution of tracers may be more trouble than it is worth, as it contains selection-function-induced features that can bias or distort the measured gradients of the distribution function. Second, as the true spatial density of the kinematic tracers is typically much smoother than the observed spatial density (as well as the selection function), it can be effectively modeled by a neural network. Except for substructures like stellar streams or moving groups, the conditional velocity distribution $p\left(\vec{v}\mid\vec{x}\right)$ is also typically smooth, making it easier to accurately capture with a conditional normalizing flow. Finally, spatial information from the observed distribution of kinematic tracers, $n_{\mathrm{obs}}\left(\vec{x}\right)$, can be reintroduced in a manner that is less likely to contaminate the spatial gradients in the CBE. Specifically, one can add in an additional term $\langle \left[\ln n\left(x\right) - \ln n_{\mathrm{obs}}\left(\vec{x}\right)\right]^2 \rangle$ to the loss function that encourages the modeled $n\left(\vec{x}\right)$ to be similar to the observed spatial density. This can also serve as a way to further decouple $\Phi(\vec x)$ from $n(\vec x)$ in configurations where they are more degenerate with one another (See Appendix~\ref{sec:uniqueness} for a discussion of the conditions under which $n$ and $\Phi$ can simultaneously be uniquely determined). The loss function then reads as
\begin{align}
  L_{\mathrm{conditional}} &=
    \Big\langle
      \sinh^{-1}\!\left(\alpha\left|\partial_t \ln f\right|\right)
      \twocolswitch{\notag \\ & \hspace{1.0em}}{}
      +
      \lambda \sinh^{-1}\!\left(
        \beta\,\mathrm{max}\left\{-\nabla^2\Phi,0\right\}
      \right)
      \twocolswitch{\notag \\ & \hspace{1.0em}}{}
      +
      \gamma_n \left(\ln n - \ln n_{\mathrm{obs}}\right)^2
    \Big\rangle_{\vec{x},\vec{v} \sim f_{\mathrm{obs}}}
  \, .
  \label{eqn:loss_conditional}
\end{align}
If an initial (ideally, unbiased) estimate of $S\left(\vec{x}\right)$ is available, then in the above loss, $n_{\mathrm{obs}}\left(\vec{x}\right)$ can be replaced by a normalizing flow model of the true spatial density $n\left(\vec{x}\right)$, learned from weighted samples of the observed positions of the tracers. In this work, we set $\gamma_n = 0$.

Table~\ref{tab:methods} summarizes the three variants of Deep Potential, indicating which functions are learned from the observed kinematic tracers and which are learned from the CBE.

\section{Test case: Mock galaxy with three-dimensional dust distribution}
\label{sec:test-case}

Here, we compare the effectiveness of the three variants of Deep Potential sketched out above (``vanilla,'' ``selection function,'' and ``conditional'') on a mock dataset. All of our code and trained models are publicly available under a permissive license that allows reuse and modification with attribution in archived form at \url{https://doi.org/10.5281/zenodo.19234304}.

\subsection{Data generation}

\begin{figure*}[t]
  \centering
  \includegraphics[width=\textwidth]{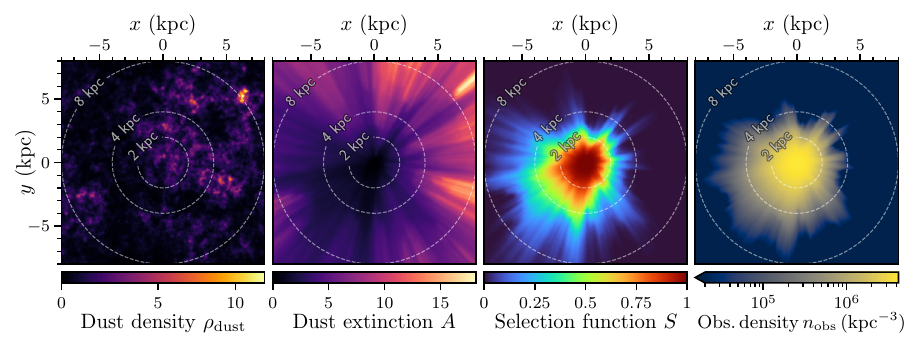}
  \caption{Two-dimensional slices (the $z = 0$ plane) through the three-dimensional dust extinction density (left) and integrated extinction (as viewed from the origin; middle left), the resulting selection function (middle right), and the expected observed spatial distribution of stars (right). The right panel shows the smooth expected density; the mock dataset used in this work is a single realization of this distribution. The observer is assumed to be at the origin. Dust extinction imprints fine angular structure on the selection function and the observed spatial distribution of stars, which can be difficult to capture with neural networks.}
  \label{fig:mock_data}
\end{figure*}

We generate mock data based on a Plummer sphere model \citep{Plummer1911DistributionGlobularClusterStars} with a complex three-dimensional distribution of dust. The Plummer sphere potential depends on radius $r$ as
\begin{align}
  \Phi\left(r\right) &=
    -\frac{G M_0}{\left(r^2 + a^2\right)^{\frac{1}{2}}}
  \, ,
\end{align}
where $G = 1$ is the gravitational constant, $M_0$ is the total mass of the system, and $a$ is a scale radius. We choose $a = 5\,\mathrm{kpc}$ and set $M_0$ such that the density at $r = 0$ is unity (in arbitrary mass units per $\mathrm{kpc}^3$). The Plummer potential admits a stationary distribution function that depends only on specific energy $E = \frac{1}{2}v^2 + \Phi$:
\begin{align}
  f\left(\vec{x},\vec{v}\right) &\propto
    \begin{cases}
      \left(-E\right)^{\frac{7}{2}} \, , & E < 0 \\
      0 \, , & E \geq 0
    \end{cases}
  \, .
\end{align}
We generate a three-dimensional distribution of dust by modeling the logarithm of the extinction density, $\ln\!\left[ \rho_{\mathrm{dust}} / \left(\mathrm{mag}\,\mathrm{kpc}^{-1}\right) \right]$, as a Gaussian process with power spectrum $P\left(k\right) \propto k^{-3}$ (with a high-frequency cutoff at $k \approx 4\,\mathrm{kpc}^{-1}$), a mean of $-0.5$, and unit variance. The left panels of Fig.~\ref{fig:mock_data} show a two-dimensional slice through the dust extinction density and the integrated extinction. We draw stars from the distribution function, subject to the constraint $r < 8\,\mathrm{kpc}$, and assign each star an absolute magnitude $M$ drawn from a normal distribution with mean $\bar{M} = 0.5\,\mathrm{mag}$ and standard deviation $\sigma_M = 2\,\mathrm{mag}$. We place the observer at $r = 0$, and calculate the apparent magnitude of each star, based on its absolute magnitude, distance modulus $\mu$ and extinction $A$. We impose an apparent magnitude cut of $m_{\mathrm{lim}} = 15$. The implied selection function is then a function of distance modulus $\mu$ and extinction $A$:
\begin{align}
\label{eqn:selfn_formula}
  S\left(\mu, A\right)
  &=
    \int_{-\infty}^{m_{\mathrm{lim}}-\mu-A}
      \hspace{-3.5em}
      p\left(M \mid \bar{M}, \sigma_M \right)
      \mathrm{d}M
  \twocolswitch{\notag \\ &=}{=}
    \frac{1}{2} \left[
      1 + \mathrm{erf} \left(
        \frac{
          m_{\mathrm{lim}} \! - \! \bar{M} \! - \! \mu \! - \! A
        }{
          \sqrt{2} \, \sigma_M
        }
      \right)
    \right]
  \, .
\end{align}
We repeatedly draw stars until we obtain $2^{22}$ ($\sim$4.19 million) observed stars. The right panels of Fig.~\ref{fig:mock_data} show a two-dimensional slice through the selection function and a projection of the expected distribution of observed stars. Both are finely structured due to dust extinction, illustrating the difficulty of capturing these structures with neural networks.

\subsection{Results with three approaches}

\begin{figure*}[t]
  \centering
  \includegraphics[width=0.9\textwidth]{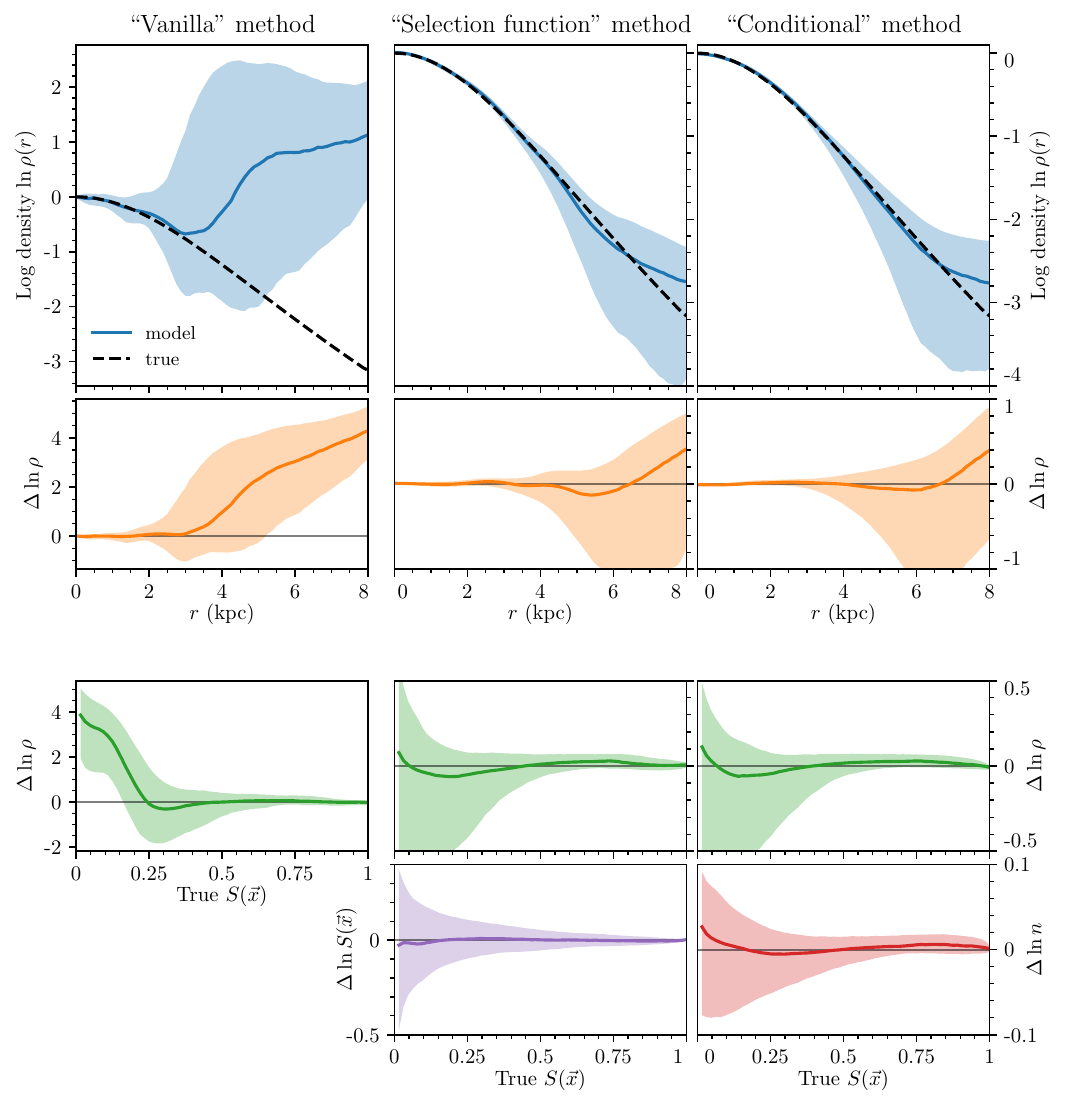}
  \caption{The gravitational densities ($\rho = \nabla^2 \Phi / 4 \pi G$) and their residuals (vs. the true densities), as recovered by the three variants of Deep Potential. In all panels, the median value of the gravitational density is plotted, along with the shaded envelopes enclosing the 16th and 84th percentiles. It should be emphasized that the envelopes do not represent statistical uncertainty, but rather angular variation at surfaces of constant $r$ or $S(\vec x)$. The percentiles are calculated over surfaces of constant distance from the origin, $r$ (top two rows, subject to a cut on the true selection function: $S(\vec x) > 0.01$) or $S(\vec x)$ (bottom two rows). \textbf{Top row:} $\ln \rho$ as a function of $r$. The dotted curves show the true density. \textbf{Second row:} Residuals $\Delta \ln \rho = \ln \rho_{\mathrm{model}} - \ln \rho_{\mathrm{true}}$ as a function of $r$. \textbf{Third row:} Residuals as a function of the true selection function $S\left(\vec{x}\right)$, illustrating the deterioration in the recovered density as the fraction of stars observed decreases. \textbf{Bottom row:} Residuals in the selection function $S\left(\vec{x}\right)$ (as recovered by the ``selection function'' method) and true spatial density of tracers $n\left(\vec{x}\right)$ (as recovered by the ``conditional'' method), each as a function of the true $S\left(\vec{x}\right)$. In the top three rows, the vertical axes are the same for the ``selection function'' and ``conditional'' methods.}
  \label{fig:model_marginals}
\end{figure*}

\begin{figure*}[t]
  \centering
  \includegraphics[width=0.9\textwidth]{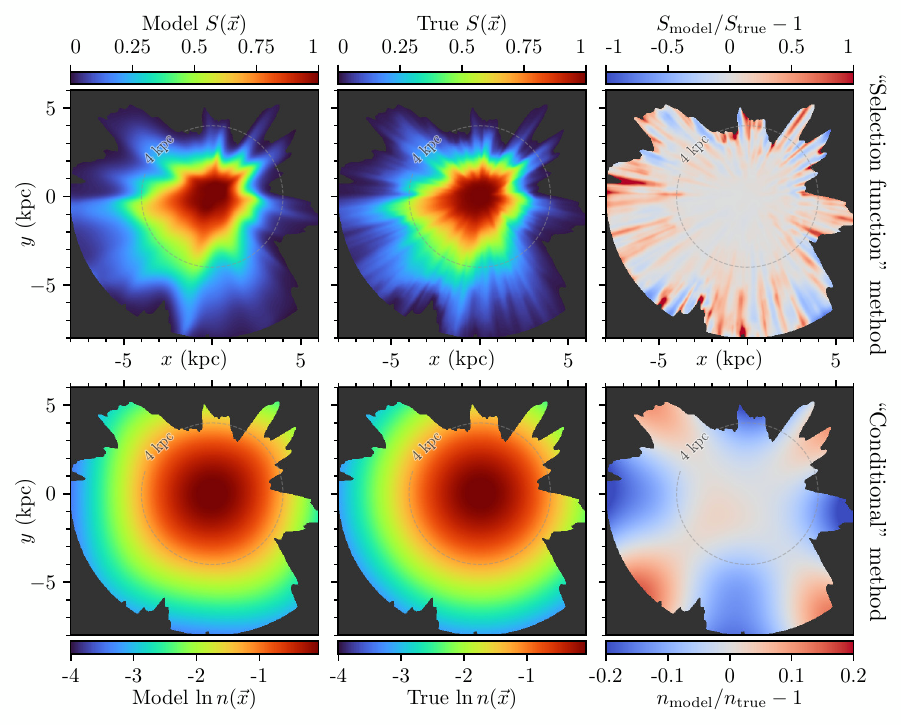}
  \caption{Recovery of the selection function and true spatial density of tracers by the ``selection function'' and ``conditional'' methods, respectively, in a two-dimensional slice through the $x-y$ plane (at $z=0$). \textbf{Top row:} The recovered selection function (left), true selection function (middle), and residuals (right) for the ``selection function'' method. \textbf{Bottom row:} The recovered true spatial density of tracers (left), true spatial density of tracers (middle), and residuals (right) for the ``conditional'' method. Dust imprints fine angular structure on the selection function, necessitating much larger models ($\sim$1.4~million parameters in this case) than are needed to capture the much smoother true spatial density of tracers ($\sim$47~thousand parameters). As seen in the top-right panel, the selection-function residuals appear as long radial spurs, illustrating the difficulty of recovering the angular structures in $S\left(\vec{x}\right)$.}
  \label{fig:selfn_n}
\end{figure*}

We run all three variants of Deep Potential on our mock dataset. We train on 75\% of the data and split off the remaining 25\% as a validation set. For each variant, we use an automatic Bayesian hyperparameter optimizer, \texttt{optuna} \citep{optuna_2019}, to determine the model architecture and training procedure that minimize the unregularized validation losses (both when learning the distribution function and when learning the potential). We train continuous normalizing flows using Conditional Flow Matching \citep{Lipman2022FlowMatching, Tong2023CFM}. We model all fields learned using the CBE and the neural networks used by Flow Matching as ResNets \citep{He2016DeepResidualLearning}. The details of our model architectures and training procedures are described in Appendix~\ref{sec:nn_details}.

For each method, we obtain a model of the gravitational potential field $\Phi\left(\vec{x}\right)$, from which we can calculate the gravitational density field $\rho\left(\vec{x}\right) = \nabla^2 \Phi / 4 \pi G$. For the ``selection function'' method, we also obtain a model of the selection function $S\left(\vec{x}\right)$, and for the ``conditional'' method, we obtain a model of the true spatial density of tracers $n\left(\vec{x}\right)$, which includes the unobserved tracers (\textit{e.g.}, those obscured by dust).

Fig.~\ref{fig:model_marginals} shows the recovered gravitational densities and their residuals (vs. the true density), both as a function of radius $r$ and the true selection function $S\left(\vec{x}\right)$. The ``vanilla'' method, which attempts to recover the true selection function by upweighting each observed tracer by $1 / S\left(\vec{x}\right)$, struggles to recover the density as the number of observed tracers falls off beyond $r \gtrsim 2\,\mathrm{kpc}$ or for $S\left(\vec{x}\right) \lesssim 0.5$. Both the ``selection function'' and ``conditional'' methods are able to accurately recover the gravitational density out to $r \sim 4\,\mathrm{kpc}$. The ``selection function'' method recovers $S\left(\vec{x}\right)$ to within $\sim$10\% accuracy for $S\left(\vec{x}\right) \gtrsim 0.25$, while the ``conditional'' method recovers $n\left(\vec{x}\right)$ to within $\sim$10\% accuracy for all $S\left(\vec{x}\right)$. Fig.~\ref{fig:selfn_n} shows the recovery of $S\left(\vec{x}\right)$ and $n\left(\vec{x}\right)$ by the ``selection function'' and ``conditional'' methods, respectively, in the $z = 0$ plane. Because $S\left(\vec{x}\right)$ contains fine angular structure, a far larger network ($\sim$1.4~million parameters) is required to model it accurately than is required for $n\left(\vec{x}\right)$ ($\sim$47~thousand parameters). The residuals in $S\left(\vec{x}\right)$ appear as long radial spurs, illustrating the difficulty of recovering the angular structures imprinted by dust extinction. This difficulty is 
compounded by the known spectral bias of deep neural networks, 
which tend to learn low-frequency functions more readily than 
high-frequency ones \citep{Rahaman2018}, though techniques such 
as Fourier feature embeddings can mitigate this 
\citep{Tancik2020}. In contrast, the fractional residuals in $n\left(\vec{x}\right)$ are smaller and smoother in $\vec{x}$. The difficulty of accurately capturing the angular structure in $S\left(\vec{x}\right)$ and $n_{\mathrm{obs}}\left(\vec{x}\right)$ is further illustrated in Appendix~\ref{sec:additional_field_figures}.

Unsurprisingly, all three methods deteriorate when the observed tracers become too sparse, as can be seen in the third row of Fig.~\ref{fig:model_marginals}, with the ``vanilla'' method deteriorating the most rapidly. Given that the ``conditional'' method requires orders of magnitude fewer parameters to achieve the same accuracy as the ``selection function'' method, we expect it to outperform the latter on real data with even more spatially complex selection functions.

A result from Jeans modeling, a closely related method to conditional Deep Potential, is that in systems with isotropic velocity distributions (such as the Plummer sphere), the 2$^\mathrm{nd}$-order velocity moments are insufficient to simultaneously constrain both the gravitational potential $\Phi\left(\vec{x}\right)$ and number density profile $n\left(\vec{x}\right)$ (See Appendix~\ref{sec:uniqueness} for a derivation of this degeneracy). The fact that we are able to simultaneously reconstruct these two fields indicates that our learned model of $p\left(\vec{v}\mid\vec{x}\right)$ accurately captures information about higher-order moments of the velocity distribution, and leverages this information to break the degeneracy between $n\left(\vec{x}\right)$ and $\Phi\left(\vec{x}\right)$. The velocity distribution in the Milky Way disk is anisotropic and has less spatial symmetry than that of our simple Plummer toy model, and should therefore contain even more information to break the potential -- tracer-density degeneracy.

We further validate this finding on a dust-free Plummer sphere in Appendix~\ref{sec:no_dust}, which isolates the method's performance from complications introduced by dust extinction and confirms that the conditional method can simultaneously recover $n(\vec x)$ and $\Phi(\vec x)$ in the absence of spatial selection effects from dust.

\section{Conclusions}
\label{sec:conclusions}

We have demonstrated a new variant of the ``Deep Potential'' approach to determining the gravitational potential of a system (such as the Milky Way) from a frozen snapshot of kinematic tracers (\textit{e.g.}, stars), which does not require modeling of finely structured spatial selection functions. Instead of modeling the full six-dimensional distribution function of the observed tracers, $f_{\mathrm{obs}}\left(\vec{x},\vec{v}\right)$, we model the conditional velocity distribution $p\left(\vec{v}\mid\vec{x}\right)$, which is unaffected by a purely spatial selection function. We simultaneously learn the gravitational potential $\Phi\left(\vec{x}\right)$ and the underlying spatial density of the entire tracer population $n\left(\vec{x}\right)$ using the collisionless Boltzmann equation under the stationarity assumption. The advantage of this method is that all of the quantities we model -- $p\left(\vec{v}\mid\vec{x}\right)$, $\Phi\left(\vec{x}\right)$, and $n\left(\vec{x}\right)$ -- typically vary smoothly in position and velocity.

We have shown that both the ``conditional'' and ``selection function'' variants of the Deep Potential method are able to accurately recover the gravitational potential in a mock dataset with a complex three-dimensional dust distribution that imprints fine angular structure on the selection function. In our mock data, the dust distribution has structure down to a scale of $\sim$0.25\,kpc, which corresponds to an angular scale of 15\,deg at a distance of 1\,kpc, or 1.8\,deg at 8\,kpc. In the real world, dust clouds have structure all the way down to sub-parsec scales. As the selection function becomes more complex, we expect the ``conditional'' method to outperform the ``selection function'' method, as the latter will struggle to accurately capture fine angular structure in both the selection function and in $\ln n_{\mathrm{obs}}\left(\vec{x}\right)$. The ``conditional'' method does not require learning either of these fields. Already at the resolution of our mock three-dimensional dust map, the model representing $\ln S\left(\vec{x}\right)$ requires $\sim$1.4~million parameters. In contrast, the model representing the full underlying tracer population, $\ln n\left(\vec{x}\right)$, which is unaffected by selection effects, requires just $\sim$47~thousand parameters (see Table~\ref{tab:hyperparams}). This discrepancy will become larger as the three-dimensional dust distribution becomes more finely structured. For this reason, we suggest that the ``conditional'' method is a promising approach for recovering the gravitational potential from real \textit{Gaia} data.

This ``conditional'' Deep Potential method is applicable not only to \textit{Gaia} data, but also to spectroscopic surveys with even more complex selection functions or uneven sky coverage, such as SDSS-V. The key requirement is that $p\left(\vec{v}\mid\vec{x}\right)$ vary smoothly enough with position $\vec{x}$ that the conditional normalizing flow can effectively interpolate between patchily observed regions of the Galaxy.

\begin{acknowledgments}
  GG and TK are supported by a Sofja Kovalevskaja Award to GG from the Alexander von Humboldt Foundation.

  Computations were performed on the HPC system Vera at the Max Planck Computing and Data Facility.
\end{acknowledgments}

\begin{contribution}
  TK originated the idea of using the conditional distribution function $p\left(\vec{v}\mid \vec{x}\right)$ to determine the gravitational potential, derived the mathematical formalism, wrote the code, ran Deep Potential on the test cases, generated most of the figures, and wrote appendices \ref{sec:nn_details}, \ref{sec:additional_field_figures}, and \ref{sec:no_dust}. GG created the mock data for the test cases, generated Fig.~\ref{fig:mock_data}, wrote most of the initial draft of the manuscript and appendix \ref{sec:uniqueness}, and provided advice and feedback on the method, code, and figures.
\end{contribution}

This work makes use of the following software packages: \texttt{astropy} \citep{astropy:2013,astropy:2018,astropy:2022}, \texttt{JAX} \citep{jax2018github}, \texttt{flowjax} \citep{flowjax25}, \texttt{diffrax} \citep{kidger2021on}, \texttt{optax} \citep{deepmind2020jax},
\texttt{equinox} \citep{kidger2021equinox}, \texttt{e3nn-jax} \citep{e3nn_paper}, \texttt{POT} \citep{POT}, \texttt{matplotlib} \citep{Hunter:2007}, \texttt{numpy} \citep{numpy}, \texttt{optuna} \citep{optuna_2019}, \texttt{python} \citep{python}, \texttt{scipy} \citep{2020SciPy-NMeth}, and \texttt{h5py} \citep{collette_python_hdf5_2014,h5py_4250762}.

Software citation information aggregated using \texttt{\href{https://www.tomwagg.com/software-citation-station/}{The Software Citation Station}} \citep{software-citation-station-paper, software-citation-station-zenodo}.

\appendix

\section{Model architectures and training procedures}
\label{sec:nn_details}

We implement all models using the \texttt{JAX} framework \citep{jax2018github}. The hyperparameters for the neural network architectures and the training procedures, determined via the \texttt{optuna} optimization framework \citep{optuna_2019}, are summarized in Table~\ref{tab:hyperparams}.

All of the scalar fields and normalizing flow vector fields are implemented as Residual Networks (ResNets). Each block returns $x + f(x)$, where $f(x)$ consists of a linear layer, followed by a $\mathrm{SiLU}(x)=x/(1+\exp(-x))$ activation function \citep{Hendrycks2016} and another linear layer. All of the models are optimized using the \texttt{AdamW} optimizer. We employ a cosine annealing learning rate schedule, which decays the learning rate from the initial value specified in Table~\ref{tab:hyperparams} to close to zero over the course of training.

\subsection{Normalizing flows}

We employ Continuous Normalizing Flows (CNFs; see Section~4 of \citealt{Papamakarios2019NormalizingFlowsReview}) to model the probability densities $n_{\mathrm{obs}}(\vec x)$ and $p(\vec v \mid \vec x)$. Unlike discrete flows that construct a transformation via a sequence of discrete layers, CNFs define the mapping from a base distribution (usually a unit Gaussian, $p_0(\vec z)$) to the target distribution ($p_1(\vec z)$) via an Ordinary Differential Equation (ODE), $\dv*{\vec z}{t} = \vec u_\theta (\vec z, t)$, where $\vec u_\theta$ is a learnable vector field parameterized by a neural network with weights $\theta$, and $t$ is an integration variable (not to be confused with the real time of any physical system). The log-probability of the target distribution is defined by the transport equation
\begin{equation}
  \ln p_1(\vec z) = \ln p_0(\vec z) - \int_{t=0}^{t=1}\dd{t}\nabla \cdot \vec u_t(\vec z, t).
\end{equation}
A significant downside of CNFs is the computational cost of integrating the ODE during inference, as it requires the use of auto-differentiable integrators to compute the gradients in the log-density.

To train the CNF, we employ Conditional Flow Matching \citep[CFM;][]{Lipman2022FlowMatching, Tong2023CFM}, a recent method that is gaining traction, largely due to its ability to train the vector field in an inference-free manner, which greatly speeds up training and facilitates the use of more complex models. The method regresses the trainable vector field $\vec u_\theta(\vec z, \vec t)$ onto a target vector field that draws straight paths between samples from the base distribution (noise), and the samples drawn from the data distribution. This avoids the costly ODE integration required during training in maximum-likelihood approaches such as FFJORD \citep{Grathwohl2018FFJORD}.

As the choice of vector field to transform from the base to the target distribution is highly degenerate with $\theta$, different forms of regularization were explored to straighten the flow trajectories and reduce numerical stiffness during inference. Perhaps the most promising was the use of minibatch optimal transport (minibatch-OT), which modifies the Conditional Flow Matching sampling scheme to respect the optimal transport map between the base and target distributions. However, minibatch-OT does not trivially extend to conditional normalizing flows \citep{Cheng2025TheCurseOfConditions}, and hence was only employed for the training of the observed spatial density $n_\mathrm{obs}(\vec x)$, not the conditional velocity distribution $p(\vec v\mid\vec x)$. Since the full optimal transport scheme is computationally expensive (the Hungarian algorithm scales with $\mathcal{O}(N^3)$), we perform it in minibatches of size 256. In practice, 512 epochs' worth of minibatch OT pairings were precomputed and later used via a dataloader during the training of the vector field. We found 512 to be a sufficient threshold, as the sampling time provides an extra layer of stochasticity, provided that after 512 training epochs, the pairings were reused and reshuffled.

We additionally employed Jacobian regularization, which punishes the rate of change of volume in $t$, by penalizing the Frobenius norm of $\pdv*{u_\theta(\vec z_i,t)}{z_j}$ \citep{Finlay2020KineticAndJacobianRegularization}. However, \texttt{optuna} decided not to make use of Jacobian regularization, likely due to its high computational cost and marginal gains.

Before feeding $t$ into the vector field network, we embed it using a sinusoidal embedding of dimension \texttt{time\_embedding\_dim} \citep{Vaswani2017}, then concatenate it with the other inputs. The frequencies are spaced geometrically between 1 and 1024. Conditioning parameters, such as the velocity, were handled by linearly appending them to the input; however, more sophisticated methods exist \citep{Perez2017FiLM}.

To summarize, the CFM loss function is computed by sampling $t\in[0, 1]$ (we let $t$ be the square root of a uniform random variable on $[0, 1]$), and sampling $z_0, z_1$ from the base and target distributions, respectively. For standard flow matching, $z_0$ and $z_1$ are drawn independently: $z_0\sim p_0(\vec z)$, $z_1\sim p_1(\vec z)$. For minibatch-OT, the pair $(z_0, z_1)$ is drawn jointly from a coupling $\pi$ that approximates the optimal transport plan between $p_0$ and $p_1$ within each minibatch. As noted above, minibatch-OT was only used for the unconditional spatial density $n_{\mathrm{obs}}(\vec x)$. For the conditional velocity distribution $p(\vec v \mid \vec x)$, we use standard flow matching (\textit{i.e.}, for a randomly drawn $\vec x_i, \vec v_i$ drawn from the training data, $z_1=\vec v_i$ is matched with $z_0$ drawn from a three-dimensional unit Gaussian, and $\vec x_i$ is provided as a conditioning input to the vector field network).
The loss for one draw is given by
\begin{align}
  \mathcal L_\mathrm{CFM}
    &= ||u_\theta (t, z_t) - u_t(z_t)||^2
       \notag \\
       &+ \lambda_\mathrm{Jacobian} \left|\left|\pdv*{u(\vec z_i,t)}{z_j}\right|\right|_F^2
    \, , \\
  z_t &= tz_1 + (1-t)z_0 \, , \\
  u_t &= z_1 - z_0 \, .
\end{align}

For all three methods, we model the full six-dimensional distribution function using the decomposition $f_{\mathrm{obs}}(\vec x, \vec v) = n_{\mathrm{obs}}(\vec x) p(\vec v \mid \vec x)$, where both distributions are represented as continuous normalizing flows. For the ``vanilla'' method, as the spatial density $n_{\mathrm{obs}}(\vec x)$ should represent an estimate of the true density, we must weight the samples by the inverse of the selection function. As $S\left(\vec{x}\right)$ approaches zero, the weights will diverge. To avoid this, we clip the weights to be $\leq 10$. For the ``selection function'' and ``conditional'' methods, we do not need to weight the samples, as in the former case, the selection function is learned from the CBE, while in the latter case, we only learn the conditional velocity distribution $p(\vec v \mid \vec x)$ from the samples and later learn the true spatial distribution of the tracers $n\left(\vec{x}\right)$ from the CBE.

\subsection{Scalar function networks}

The gravitational potential $\Phi(\vec{x})$, the selection function correction $\ln S(\vec{x})$, and the true number density $\ln n(\vec{x})$ are all scalar functions that take 3-vectors (representing $\vec{x}$) as inputs. 
To assist the networks in capturing the complex angular structure of the selection function (as visualized in Figs.~\ref{fig:mock_data} and \ref{fig:selfn_mw_comparison}), we augment the input coordinates for the $\ln S(\vec{x})$ and $n_{\mathrm{obs}}(\vec{x})$ networks. In addition to Cartesian coordinates, we compute real Spherical Harmonics $Y_{l}^{m}(\theta, \phi)$ up to a maximum degree of \texttt{SH\_embedding\_lmax}. These harmonic features are concatenated with the input vector, allowing the network to explicitly leverage angular basis functions suitable for modeling extinction features on the sky. We found that while higher degrees of spherical harmonics help with training, values $\geq 10$ increase numerical stiffness and make the inference unstable. Additionally, perhaps unsurprisingly, spherical harmonics were not utilized for the learning of $\ln n(\vec x)$ because it is less sensitive to angular features arising from extinction.

We use \texttt{optuna} to perform a Bayesian search over the hyperparameter space. For the normalizing flows, the objective is to maximize the log-likelihood of the validation set. For the CBE-based networks ($\Phi$, $S$, $n$), the objective is to minimize the total physics loss (stationarity penalties plus regularization terms; see Eqs.~\ref{eqn:loss_vanilla}, \ref{eqn:loss_selfn} and \ref{eqn:loss_conditional}) on the validation set.

\subsection{Computational costs}

Compared to \cite{KaldaGreen2025DeepPotentialMW1kpc}, the performance of the normalizing flows has undergone dramatic improvements. From most to least important, the improvements arose from switching from FFJORD to Conditional Flow Matching, using ResNets and positional embeddings (instead of Multilayer Perceptrons), transitioning from \texttt{TensorFlow}~2 \citep{tensorflow2015-whitepaper} to \texttt{JAX}, and using the SiLU activation (instead of $\tanh$). The speed-up was observed to be larger than a factor of 100 for more complex flows. This speed-up allowed us to use \texttt{optuna} (requiring many trials) to optimize hyperparameters. The computational times of different components of the model on an Nvidia A100 are listed in Table~\ref{tab:hyperparams}.

\begin{table*}[h]
  \centering
  \hspace{-70pt}
  \begin{tabular}{
    >{\centering}m{0.1\textwidth} 
    m{10em}
    >{\centering}m{0.11\textwidth}|
    >{\centering}m{0.185\textwidth}|
    >{\centering\arraybackslash}m{0.1\textwidth}
  }
    \hline
    \hline

    & &\multicolumn{3}{c}{\textbf{Method}}\\
    \cline{3-5}
    & \textbf{Parameter} & Vanilla & Selection function & Conditional \\
    \cline{2-5}

    \multirow{4}{*}{{\textbf{General}}} 
    & Base Network & \multicolumn{3}{c}{ResNet} \\ \cline{2-2}
    & Activation & \multicolumn{3}{c}{SiLU} \\ \cline{2-2}
    & Optimizer & \multicolumn{3}{c}{AdamW} \\ \cline{2-2}
    & LR Scheduler & \multicolumn{3}{c}{Cosine annealing}\\
    \hline

    \multirow{2}{*}{{\parbox{2cm}{\centering \textbf{Minibatch-OT}}}} 
    & \texttt{epochs} & \multicolumn{3}{c}{512} \\ \cline{2-2}
    & \texttt{batch\_size} & \multicolumn{3}{c}{256} \\
    \hline

    \multirow{7}{*}{{\boldmath$n_{\mathrm{obs}}(\vec{x})$}} 
    & \texttt{width} $\times$ \texttt{depth} & \multicolumn{3}{c}{$115\times 6$} \\ \cline{2-2}
    & \texttt{time\_embedding\_dim} & \multicolumn{3}{c}{$50$} \\ \cline{2-2}
    & \texttt{SH\_embedding\_lmax} & \multicolumn{3}{c}{$3$} \\ \cline{2-2}
    & \texttt{learning\_rate} & \multicolumn{3}{c}{$2.0 \times 10^{-3}$} \\ \cline{2-2}
    & \texttt{epochs} & \multicolumn{3}{c}{$1558$} \\ \cline{2-2}
    & \# of parameters & \multicolumn{3}{c}{\num{166656}} \\ \cline{2-2}
    & Training time & \multicolumn{3}{c}{\SI{47}{min} training + \SI{90}{min} sampling} \\
    \hline

    \multirow{6}{*}{{\boldmath$p(\vec{v}\mid\vec{x})$}} 
    & \texttt{width} $\times$ \texttt{depth} & \multicolumn{3}{c}{$57 \times 10$} \\ \cline{2-2}
    & \texttt{time\_embedding\_dim} & \multicolumn{3}{c}{$50$} \\ \cline{2-2}
    & \texttt{learning\_rate} & \multicolumn{3}{c}{$1.16 \times 10^{-3}$} \\ \cline{2-2}
    & \texttt{epochs} & \multicolumn{3}{c}{$852$} \\ \cline{2-2}
    & \# of parameters & \multicolumn{3}{c}{\num{69580}} \\ \cline{2-2}
    & Training time & \multicolumn{3}{c}{\SI{23}{min} training + \SI{45}{min} sampling} \\
    \hline

    \multirow{6}{*}{{\boldmath$\Phi(\vec{x})$}} 
    & \texttt{width} $\times$ \texttt{depth} & $58 \times 6$ & $29 \times 8$ & $29 \times 8$ \\ \cline{2-2}
    & $\ell_2$ regularization & $0.23$ & $0.86$ & $0.31$ \\ \cline{2-2}
    & \texttt{learning\_rate} & $5.7 \times 10^{-3}$ & $7.9 \times 10^{-3}$ & $8.4 \times 10^{-3}$ \\ \cline{2-2}
    & \texttt{epochs} & $155$ & $116$ & $116$ \\ \cline{2-2}
    & \# of parameters & \num{42054} & \num{14533} & \num{14533} \\ \cline{2-2}
    & Training time & \SI{12}{min} & -- & -- \\
    \hline

    \multirow{5}{*}{{\boldmath$S(\vec{x}),n(\vec x)$}} 
    & \texttt{width} $\times$ \texttt{depth} & -- & $240 \times 12$ & $75 \times 4$ \\ \cline{2-2}
    & \texttt{SH\_embedding\_lmax} & -- & $3$ & $0$ \\ \cline{2-2}
    & $\ell_2$ regularization & -- & $0.73$ & $0.16$ \\ \cline{2-2}
    & \# of parameters & -- & \num{1398970} & \num{46585} \\ \cline{2-2}
    & Training time & -- & \SI{20}{min} & \SI{12}{min} \\
    \hline
  \end{tabular}
  \caption{Hyperparameters of the different ``Deep Potential'' methods used in this study. The hyperparameters were tuned using the \texttt{optuna} Python package. The training time benchmarks were computed on an Nvidia A100 GPU. The $n_{\mathrm{obs}}(\vec x)$ and $p(\vec v \mid \vec x)$ flows were trained in common for all three methods, while the $\Phi(\vec x)$ networks were trained separately for each method. For the ``selection function'' and ``conditional'' methods, we used the same network structure for $\Phi$, determined by \texttt{optuna} (based on the ``selection function'' method), but allowed \texttt{optuna} to determine a separate learning rate and regularization strength for the ``conditional'' case. The $S(\vec x)$ and $n(\vec x)$ networks were only trained for the ``selection function'' and ``conditional'' methods, respectively. These last two networks were trained simultaneously with the potential, and thus share learning rates and training times with $\Phi$.}
  \label{tab:hyperparams}
\end{table*}

\section{Recovery of selection and distribution functions}
\label{sec:additional_field_figures}

\begin{figure*}
  \centering
  \includegraphics[scale=0.9]{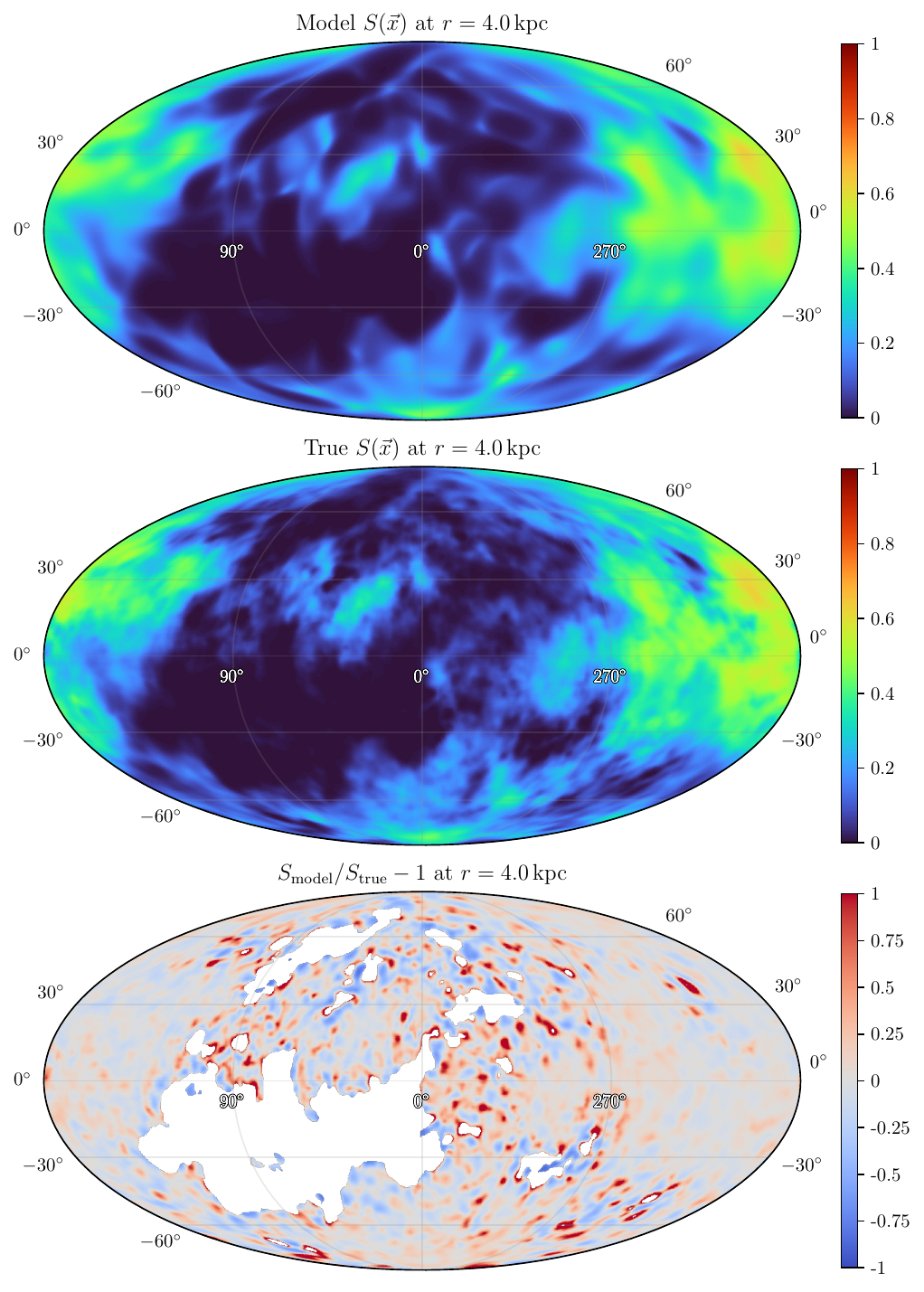}
  \caption{Sky plots of the ``selection function'' method's prediction of the selection function $S(\vec x)$, the true selection function, and their fractional residuals on the surface of a sphere of radius \SI{4}{kpc}, centered on the mock observer. Areas with $S_\mathrm{true} < 0.01$ are masked. The differences can be attributed to shot noise in areas of low completeness, and to the difficulty of capturing fine angular structure with a neural network. Nevertheless, with a mock selection function with structure down to $\sim$0.25\,kpc scales, our selection-function implementation performs well, albeit at the cost of large model sizes (here, $\sim$1.4~million parameters).}
  \label{fig:selfn_mw_comparison}
\end{figure*}

\begin{figure}
  \centering
  \includegraphics{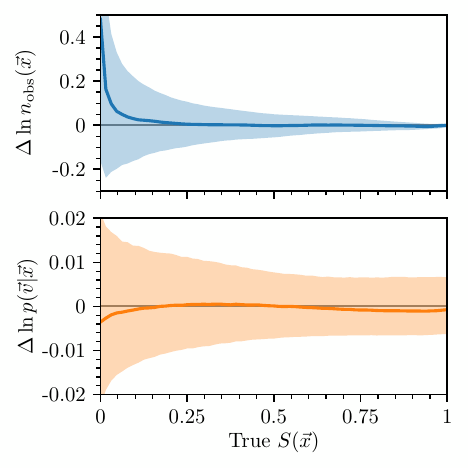}
  \caption{\textbf{Top panel:} Residuals between our normalizing flow model of the distribution of observed kinematic tracers and the ground truth: $\Delta \ln n_{\mathrm{obs}}\left(\vec{x}\right) \equiv \ln n_{\mathrm{obs}}\left(\vec{x}\right) - \ln \left[n\left(\vec{x}\right) S\left(\vec{x}\right)\right]$, where $n\left(\vec{x}\right)$ and $S\left(\vec{x}\right)$ are the true spatial density of tracers (including those that are not observed) and selection function, respectively. \textbf{Bottom panel:} Residuals between our conditional normalizing flow model of the velocity distribution and the ground truth. The conditional distribution $p\left(\vec{v}\mid\vec{x}\right)$ is recovered to within $\sim$2\%, even for $S\left(\vec{x}\right)$ as low as $\sim$0.01. Due to the complex nature of the spatial selection function, the fractional errors in the recovered $n_{\mathrm{obs}}\left(\vec{x}\right)$ are larger, and the median residual diverges substantially from zero at low $S\left(\vec{x}\right)$.}
  \label{fig:df_comparison}
\end{figure}

Here, we present additional figures illustrating the recovery of various fields by the ``selection function'' and ``conditional'' methods.

Fig.~\ref{fig:selfn_mw_comparison} shows a sky plot (at $r=\SI{4}{kpc}$) of the ``selection function'' method's recovery of the selection function in comparison with the ground truth. Dust extinction imprints fine angular patterns on the selection function. In particular, highly extincted areas have low numbers of ``detected'' mock stars, and hence recovery of the selection function is limited by shot noise.

Fig.~\ref{fig:df_comparison} compares our normalizing flow representations of $n_{\mathrm{obs}}\left(\vec{x}\right)$ and $p\left(\vec{v}\mid\vec{x}\right)$ with the ground truth. The conditional velocity distribution is recovered to within $\sim$2\% accuracy all the way down to $S\left(\vec{x}\right) \sim 0.01$. In contrast, due to the complex patterns imprinted by the spatial selection function, the residuals in the recovered $n_{\mathrm{obs}}\left(\vec{x}\right)$ are larger and display a large positive trend at low $S\left(\vec{x}\right)$. The ``conditional'' method only makes use of the conditional velocity distribution, while the ``selection function'' method additionally relies on the modeled $n_{\mathrm{obs}}\left(\vec{x}\right)$.

\section{Uniqueness of solutions}
\label{sec:uniqueness}

Conditional Deep Potential only uses information in $p\left(\vec{v}\mid\vec{x}\right)$, and treats both the kinematic tracer density $n\left(\vec{x}\right)$ and gravitational potential $\Phi\left(\vec{x}\right)$ as unknown fields to be determined. It is therefore natural to ask whether this problem is underconstrained. We approach this question by relating conditional Deep Potential to the Jeans method \citep{Jeans1915OnTheoryStarStreaming, Jeans1922MotionsOfStars, Binney1980JeansEllipticalGalaxies, Cappellari2008AnisotropicJeansModels}, which determines the potential based on empirically measured moments of the velocity distribution:
\begin{align}
  \langle v_i v_j \cdots \rangle
  \equiv
  \int v_i v_j \cdots p\left(\vec{v}\mid\vec{x}\right) \mathrm{d}^3\vec{v}
  \, .
\end{align}
The $\left(k+1\right)^\mathrm{st}$-order Jeans equation is obtained by taking the $k^\mathrm{th}$ velocity moment of the CBE (Eq.~\ref{eqn:CBE}) under the stationarity assumption ($\partial_t f = 0$), and reads
\begin{align}
  0 \, &=
  \partial_i \left( \langle v_i v_{j_1} \cdots v_{j_k} \rangle n \right)
  + k \, n \langle v_{(j_1} \cdots v_{j_{k-1}} \rangle \partial_{j_k)} \Phi
  \, ,
\end{align}
where $\partial_i \equiv \partial / \partial x_i$ and the subscript parentheses in the second term indicate symmetrization of the indices. As in conditional Deep Potential, we will analyze uniqueness of solutions arising from the above equation, assuming that the velocity moments have been measured, but that the fields $n$ and $\Phi$ are unknown. As both conditional Deep Potential and this ``conditional'' variant of the Jeans method use information on the velocity distribution to simultaneously determine the potential and kinetic tracer density under the assumption of stationarity, the conditions under which they can produce unique solutions should be highly similar.

Given the existence of a stationary solution ($\ln n$, $\Phi$), we look for additional solutions of the form $\ln n' = \ln n + \delta \ln n$ and $\Phi' = \Phi + \delta \Phi$. Plugging these forms of the solution into the $k^\mathrm{th}$-order Jeans equation yields the condition
\begin{align}
  0 \, &=
  \langle v_i v_{j_1} \cdots v_{j_k} \rangle \partial_i \! \left(\delta \ln n\right)
  + k \langle v_{(j_1} \cdots v_{j_{k-1}} \rangle \partial_{j_k)} \delta\Phi
  \, .
\end{align}
If solutions to $\delta \ln n$ and $\delta \Phi$ exist for a given set of velocity moments, then those moments do not fully constrain both $n$ and $\Phi$, and additional information, such as the observed tracer density $n$ or higher-order velocity moments, must be provided. Whether or not solutions to the Jeans equations exist depends critically on the properties of the velocity moments and how they vary spatially.

At 2$^\mathrm{nd}$ order, several classes of systems are underconstrained. Here, we will consider systems with isotropic velocity distributions at every point in space: $\langle v_i v_j \rangle = V\left(\vec{x}\right) \delta_{ij}$, where $V\left(\vec{x}\right)$ is some arbitrary function of space. The Plummer sphere toy model that we employ in Section~\ref{sec:test-case} falls into this class of systems, with $V\left(\vec{x}\right) = -\frac{1}{6} \Phi\left(\vec{x}\right)$. With 2$^\mathrm{nd}$-order moments of this form, the 2$^\mathrm{nd}$-order Jeans uniqueness condition simplifies to
\begin{align}
  0 \, &=
  V \nabla \! \left(\delta \ln n\right)
  + \nabla \left(\delta\Phi\right)
  \, .
\end{align}
This equation is solved by $\delta \ln n = g(V)$, $\delta \Phi = h(V)$, where $g\left(V\right)$ is any scalar function and $h'\left(V\right) = -V g'\left(V\right)$. Thus, if $p\left(\vec{v}\mid\vec{x}\right)$ only captures information contained in the 2$^\mathrm{nd}$-order velocity moments, the conditional Deep Potential method will fail to obtain unique solutions.

However, higher-order velocity moments do fully constrain the Plummer sphere model. Odd-order velocity moments are zero in an isotropic system, so we proceed directly to the uniqueness constraint from the 4$^\mathrm{th}$-order Jeans equation. The 4$^\mathrm{th}$ velocity moments of an isotropic system can be parameterized as $\langle v_i v_j v_k v_\ell \rangle = 3 \, \kappa\left(\vec{x}\right) V^2\left(\vec{x}\right) \delta_{(ij} \delta_{k\ell)}$, where $\kappa\left(\vec{x}\right)$ is a function proportional to the kurtosis of the individual velocity component distributions (defined so that $\kappa = 1$ for a Gaussian velocity distribution). Plugging this into the 4$^\mathrm{th}$-order Jeans equation and eliminating $\Phi$ using the 2$^\mathrm{nd}$-order Jeans result $n \partial_j \Phi = -\partial_j\left(V n\right)$ yields
\begin{align}
  0 &=
  \delta_{(jk} \partial_{\ell)} \left( \kappa V^2 n \right)
  - V \delta_{(jk} \partial_{\ell)} \left( V n \right)
  \, .
\end{align}
Taking the $j = k = \ell$ terms (or equivalently, taking the $j=k$ trace), we find that
\begin{align}
  0 &=
  \nabla \left( \kappa V^2 n \right) - V \nabla \left( V n \right)
  \, .
\end{align}
This simplifies to an equation for the gradients of $\ln n$
\begin{align}
  0 &=
  \left(\kappa - 1\right) V^2 \nabla \left( \ln n \right)
  + \nabla \left[ \left( \kappa - \frac{1}{2} \right) V^2 \right]
  \, .
\end{align}
If $\kappa = 1$ everywhere (the case for a purely Gaussian velocity distribution), then the 4$^\mathrm{th}$-order Jeans equation yields no new information on $\ln n$ in an isotropic system (though it does require $V$ to be constant in space). However, when $\kappa \neq 1$, we obtain an exact solution for the gradients $\ln n$:
\begin{align}
  \nabla \left(\ln n\right)
  &=
  -\, \frac{
    \nabla \left[ \left( \kappa - \frac{1}{2} \right) V^2 \right]
  }{
    \left(\kappa - 1\right) V^2
  }
  \, .
  \label{eqn:lnn_solution_4th_order_jeans}
\end{align}
This solution can be plugged into the 2$^\mathrm{nd}$-order Jeans equation to also uniquely determine the gradients of $\Phi$. For the Plummer sphere, one can calculate that $V = -\Phi / 6$ and $\kappa = 6/7$, which combined with Eq.~\eqref{eqn:lnn_solution_4th_order_jeans} yield the correct density distribution $n\left(r\right) \propto (r^2+1)^{-5/2}$.

Thus, in systems such as the isotropic Plummer sphere, with a high level of symmetry, non-Gaussianity encoded in higher-order moments of the velocity distribution does uniquely constrain both $\ln n$ and $\Phi$. The fact that conditional Deep Potential is able to accurately constrain these two fields simultaneously in our tests (Section~\ref{sec:test-case}) demonstrates that it is able to access information contained in 4$^\mathrm{th}$ and higher-order moments of the velocity distribution $p\left(\vec{v}\mid\vec{x}\right)$.

We expect the uniqueness constraints to be significantly less demanding in the Milky Way, as it contains far less symmetry than our Plummer sphere toy model. The Milky Way's velocity distribution is not generally isotropic, and odd-order velocity moments are generally non-zero (particularly in the disk, where there is net rotation). Information from 2$^\mathrm{nd}$ and 3$^\mathrm{rd}$-order velocity moments are therefore likely to overconstrain the gravitational potential and the density of appropriately chosen stellar tracer populations in the Milky Way.

\section{Plummer sphere with no dust}
\label{sec:no_dust}

To validate the performance of the conditional Deep Potential method without the complications introduced by dust extinction, we apply it to a Plummer sphere with no dust.
The mock data is generated from the same Plummer sphere model described in Section~\ref{sec:test-case}, with identical parameters, except that dust extinction $A$ is equal to zero for all stars. A selection function still exists due to the magnitude limit (fainter stars at larger distances are preferentially excluded) and due to the distance limit $r < \SI{8}{kpc}$. The resulting selection function is purely radial (\textit{i.e.}, it has no angular dependence). The number of stars we draw is set so that the local density of stars in the vicinity of the origin ($r = 0$) matches that of our dusty mock dataset. Because of the lack of dust extinction, our dust-free mock dataset has 3.52 times as many stars as the dusty dataset.

We train the conditional Deep Potential method using the same architecture and hyperparameter optimization procedure described in Appendix~\ref{sec:nn_details}. A minimal summary of the key parameters is provided in Table~\ref{tab:hyperparams_no_dust}.

\subsection{Results}

\begin{figure*}[t]
  \centering
  \includegraphics[width=\textwidth]{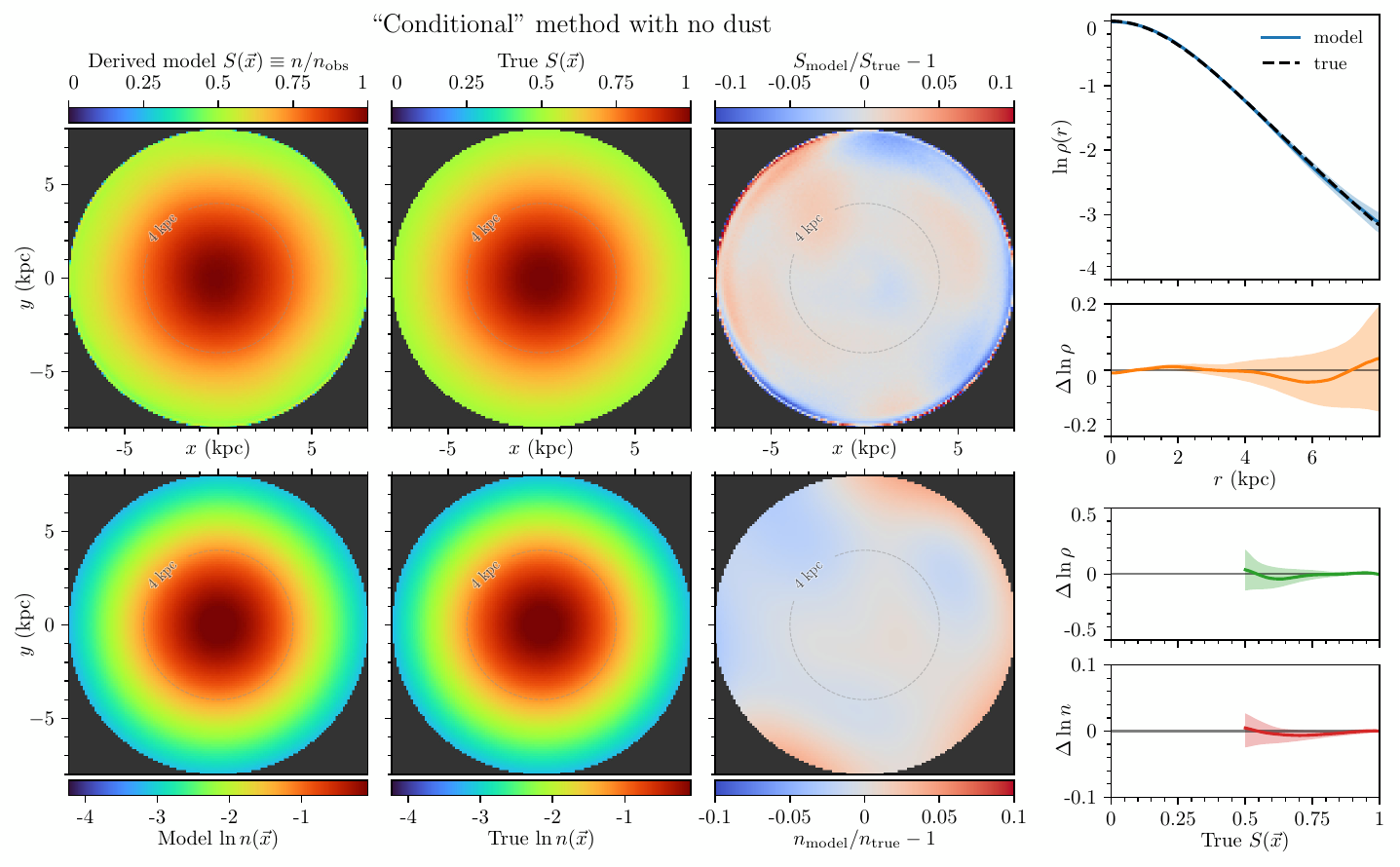}
  \caption{Performance of the conditional Deep Potential method on the dust-free Plummer sphere. \textbf{Top row, left three panels:} The derived selection function, defined as $S\left(\vec{x}\right) \equiv n\left(\vec{x}\right) / n_{\mathrm{obs}}\left(\vec{x}\right)$, the true selection function, and the fractional residuals, shown in a two-dimensional slice through the $z = 0$ plane. The derived selection function is expected to behave almost identically to the ``selection function'' method of Deep Potential. \textbf{Bottom row, left three panels:} The recovered true spatial density of tracers $\ln n\left(\vec{x}\right)$, the ground truth, and the fractional residuals. Both $S_\mathrm{model}$ and $n_\mathrm{model}$ are recovered to within $\sim$5\% throughout the volume and do not exhibit any fine angular structure, unlike for the dusty Plummer sphere. \textbf{Right column, from top to bottom:} The median gravitational density $\ln \rho = \ln(\nabla^2\Phi/(4\pi G))$ as a function of $r$; residuals $\Delta \ln \rho$ as a function of $r$; residuals $\Delta \ln \rho$ as a function of the true $S\left(\vec{x}\right)$; and residuals $\Delta \ln n$ as a function of the true $S\left(\vec{x}\right)$. The clipping in the last two plots is caused by the selection function being greater than $\sim 0.5$ throughout the volume. Shaded regions enclose the 16th and 84th percentiles. Compare to Figs.~\ref{fig:model_marginals} and \ref{fig:selfn_n}, which show results obtained for a dusty Plummer sphere.}
  \label{fig:no_dust_conditional}
\end{figure*}

Fig.~\ref{fig:no_dust_conditional} shows the results of running conditional Deep Potential on the dust-free Plummer sphere. The top-right panels show our recovered $\ln \rho$ and its residuals as a function of radius, while the bottom-right panels show residuals in $\ln \rho$ and $\ln n$ as a function of the selection function $S\left(\vec{x}\right)$.
At small radii ($r \lesssim 3\,\mathrm{kpc}$), we obtain comparable results as for the dusty mock dataset, consistent with the selection function being close to unity at short distances in both scenarios. However, the dust-free results remain highly accurate out to far larger radii (with $\lesssim 20\%$ density errors at 8~kpc). This demonstrates that the dominant source of degradation in the dusty case is the reduced effective sample size in heavily extincted regions.

Using the results of conditional Deep Potential, it is possible to estimate the selection function by comparison of the kinematically determined model $\ln n$ and the $\ln n_{\mathrm{obs}}$ learned from the observed stars: $S_{\mathrm{model}} \equiv n / n_{\mathrm{obs}}$. The top-left panels of Fig.~\ref{fig:no_dust_conditional} show the derived model selection function recovered in this manner, the true selection function, and their residuals in a two-dimensional slice through the $z = 0$ plane. In this dust-free case, the residuals in the model selection function are on the order of a few percent, except in a thin shell at the edge of the volume, where our normalizing flow model for $\ln n_{\mathrm{obs}}$ has difficulty recovering the discontinuous fall-off (to zero) in the density of observed stars.

The bottom-left panels of Fig.~\ref{fig:no_dust_conditional} show our recovered $\ln n$, the true $\ln n$, and their residuals. The modeled $n\left(\vec{x}\right)$ is accurate to within a few percent throughout the entire volume.

As discussed in Appendix~\ref{sec:uniqueness}, for the Plummer sphere's isotropic velocity distribution, the solutions to the fields $n\left(\vec{x}\right)$ and $\Phi\left(\vec{x}\right)$ are degenerate when only information on the 2$^\mathrm{nd}$-order velocity moments of $p\left(\vec{v}\mid\vec{x}\right)$ are available. The successful simultaneous recovery of both fields confirms that the conditional normalizing flow captures the non-Gaussianity of $p\left(\vec{v}\mid\vec{x}\right)$ -- specifically, its kurtosis -- and that our CBE-based optimization leverages this higher-order information to break the degeneracy between $n\left(\vec{x}\right)$ and $\Phi\left(\vec{x}\right)$.

\begin{table}[h]
  \centering
  \begin{tabular}{
    >{\centering}m{0.16\columnwidth}
    m{12em}
    >{\centering\arraybackslash}m{0.28\columnwidth}
  }
    \hline
    \hline
    & \textbf{Parameter} & \textbf{Value} \\
    \hline
    \multirow{6}{*}{{\boldmath$n_{\mathrm{obs}}(\vec{x})$}}
    & \texttt{width} $\times$ \texttt{depth} & $98\times 7$ \\ \cline{2-2}
    & \texttt{time\_embedding\_dim} & $40$ \\ \cline{2-2}
    & \texttt{SH\_embedding\_lmax} & $3$ \\ \cline{2-2}
    & \texttt{learning\_rate} & $1.3 \times 10^{-3}$ \\ \cline{2-2}
    & \texttt{epochs} & $478$ \\ \cline{2-2}
    & \# of parameters & \num{140455} \\
    \hline
    \multirow{5}{*}{{\boldmath$p(\vec{v}\mid\vec{x})$}}
    & \texttt{width} $\times$ \texttt{depth} & $109 \times 7$ \\ \cline{2-2}
    & \texttt{time\_embedding\_dim} & $42$ \\ \cline{2-2}
    & \texttt{learning\_rate} & $6.3 \times 10^{-4}$ \\ \cline{2-2}
    & \texttt{epochs} & $713$ \\ \cline{2-2}
    & \# of parameters & \num{173564} \\
    \hline
    \multirow{5}{*}{{\boldmath$\Phi(\vec{x})$}}
    & \texttt{width} $\times$ \texttt{depth} & $29 \times 8$ \\ \cline{2-2}
    & $\ell_2$ regularization & $0.31$ \\ \cline{2-2}
    & \texttt{learning\_rate} & $8.4 \times 10^{-3}$ \\ \cline{2-2}
    & \texttt{epochs} & $80$ \\ \cline{2-2}
    & \# of parameters & \num{32695} \\
    \hline
    \multirow{4}{*}{{\boldmath$n(\vec{x})$}}
    & \texttt{width} $\times$ \texttt{depth} & $75 \times 4$ \\ \cline{2-2}
    & \texttt{SH\_embedding\_lmax} & $0$ \\ \cline{2-2}
    & $\ell_2$ regularization & $0.16$ \\ \cline{2-2}
    & \# of parameters & \num{77175} \\
    \hline
  \end{tabular}
  \caption{Hyperparameters of the conditional Deep Potential method for the dust-free Plummer sphere test case. The hyperparameters were tuned using \texttt{optuna}.}
  \label{tab:hyperparams_no_dust}
\end{table}


\bibliography{bibliography}{}
\bibliographystyle{aasjournalv7}



\end{document}